\documentclass[reqno,centertags,12pt]{amsart}
\usepackage{amsmath,amsthm,amscd,amssymb}
\usepackage{latexsym}

\newcommand{\bbR}{{\mathbb{R}}}

\newcommand{\bbZ}{{\mathbb{Z}}}

\newcommand{\lb}{\label}
\newcommand{\f}{\frac}

\newcommand{\ol}{\overline}

\newcommand{\tr}{\text{\rm{Tr}}}

\newcommand{\ess}{\text{\rm{ess}}}
\newcommand{\eff}{\text{\rm{eff}}}
\newcommand{\ac}{\text{\rm{ac}}}

\newcommand{\supp}{\text{\rm{supp}}}

\newcommand{\bi}{\bibitem}

\newcommand{\beq}{\begin{equation}}
\newcommand{\eeq}{\end{equation}}
\newcommand{\ba}{\begin{align}}
\newcommand{\ea}{\end{align}}
\newcommand{\veps}{\varepsilon}

\newcounter{smalllist}
\newenvironment{SL}{\begin{list}{{\rm\roman{smalllist})}}{%
\setlength{\topsep}{0mm}\setlength{\parsep}{0mm}\setlength{\itemsep}{0mm}%
\setlength{\labelwidth}{2em}\setlength{\leftmargin}{2em}\usecounter{smalllist}%
}}{\end{list}}

\DeclareMathOperator{\Real}{Re}

\allowdisplaybreaks
\numberwithin{equation}{section}

\newtheorem{theorem}{Theorem}[section]
\newtheorem*{t1}{Theorem 1}
\newtheorem*{t2}{Theorem 2}
\newtheorem*{t3}{Theorem 3}
\newtheorem*{t4}{Theorem 4}
\newtheorem{proposition}[theorem]{Proposition}
\newtheorem{lemma}[theorem]{Lemma}
\newtheorem{corollary}[theorem]{Corollary}
\theoremstyle{definition}

\newtheorem{example}[theorem]{Example}

\theoremstyle{remark}

\newcommand{\abs}[1]{\lvert#1\rvert}

\begin{document}
\title[Bound States and the Szeg\H{o} Condition]{Bound States and the Szeg\H{o} Condition \\
for Jacobi Matrices and \\ Schr\"odinger Operators}
\author[D. Damanik, D. Hundertmark, and B. Simon]{David Damanik$^{1,2}$, Dirk Hundertmark$^1$, 
and Barry Simon$^{1,3}$}

\thanks{$^1$ Mathematics 253-37, California Institute of Technology, Pasadena, CA 91125. 
E-mail: damanik@its.caltech.edu; dirkh@its.caltech.edu; bsimon@caltech.edu}
\thanks{$^2$ Supported in part by NSF grant DMS-0227289}
\thanks{$^3$ Supported in part by NSF grants DMS-9707661, DMS-0140592}

\date{August 19, 2002}

\begin{abstract} For Jacobi matrices with $a_n =1+(-1)^n\alpha n^{-\gamma}$, $b_n=(-1)^n \beta 
n^{-\gamma}$, we study bound states and the Szeg\H{o} condition. We provide a new proof of 
Nevai's result that if $\gamma >\f12$, the Szeg\H{o} condition holds, which works also if one 
replaces $(-1)^n$ by $\cos (\mu n)$. 
We show that if $\alpha =0$, $\beta\neq 0$, and $\gamma <\f12$, the Szeg\H{o} condition fails. 
We also show that if $\gamma =1$, $\alpha$ and $\beta$ are small enough 
($\beta^2 + 8 \alpha^2 < \f{1}{24}$ will do), then the Jacobi matrix has finitely many bound 
states (for $\alpha =0$, $\beta$ large, it has infinitely many).
\end{abstract}

\maketitle

\section{Introduction} \lb{s1}

This paper focuses on Jacobi matrices, that is, operators $J$ on $\ell^2 (\bbZ_+)$, where 
$\bbZ_+ = \{1,2,\dots \}$, given by $(b_n$ real, $a_n >0$)
\begin{equation} \lb{1.1} 
(Ju)(n) =a_n u(n+1) + b_n u(n) + a_{n-1} u(n-1)
\end{equation}
where the $a_{n-1}u(n-1)$ term is dropped if $n=1$. We define $J_0$ by $a_n\equiv 1$, $b_n\equiv 0$, 
and will suppose $J-J_0$ is compact, so $\sigma_{\ess}(J) =[-2,2]$. We are interested especially in 
the Szeg\H{o} condition, 
\begin{equation} \lb{1.2} 
Z(J)\equiv \f{1}{2\pi} \int_{-2}^2 \log \biggl( \f{\sqrt{4-E^2}}{2\pi \f{d\nu_{\ac}}{dE}}\biggr) 
\f{dE}{\sqrt{4-E^2}} < \infty
\end{equation}
where $\nu$ is the spectral measure for $J$ and the vector $\delta_1$. We will also consider some 
aspects of Schr\"odinger operators $-\Delta + V$\!. 

In 1979, Nevai \cite{Nev79} proved a conjecture of Askey that if 
\begin{equation} \lb{1.3} 
a_n = 1+ \f{(-1)^n \alpha}{n} + O(n^{-2}) \qquad b_n = \f{(-1)^n\beta}{n} + O(n^{-2})
\end{equation}
then the Szeg\H{o} condition holds. Our goal here is to understand this result from the point of 
view of sum rules recently used to study the Szeg\H{o} condition by Killip-Simon \cite{KS} and 
Simon-Zlato\v{s} \cite{SZ}, and to consider various extensions and borderline cases, in particular, 
the following four questions: 
\begin{SL} 
\item[(1)] Nevai \cite{Nev79} allows replacement of $\f{(-1)^n}{n}$ by $\f{(-1)^n}{n^\gamma}$ with 
$\gamma > \f12$ and still gets \eqref{1.2}. Is $\gamma=\f12$ a borderline or just where Nevai's 
method fails? We will see that $\gamma=\f12$ is indeed a borderline and that $a_n =1$, $b_n =
\f{(-1)^n}{n^\gamma}$ obeys \eqref{1.2} if and only if $\gamma >\f12$. This is a subtle issue: 
one might think the key is that $b_{n+1}-b_n$ decay faster than $n^{-1}$, in which case $\gamma=
\f12$ is not special but, as is the case in many other situations \cite{S265}, $b_n\in\ell_2$ 
is critical; see Theorem 2 below. 

\item[(2)] What is the condition on the errors $O(n^{-2})$ in \eqref{1.3}? Nevai actually shows 
if those errors, $e_a(n), e_b(n)$, obey $\sum_{n=2}^\infty (\log n) \abs{e_{j}(n)}
<\infty$ for $j\in\{a,b\}$, then \eqref{1.2} still holds. In line with the advances in 
\cite{KS,SZ}, we will only 
require $\sum_{n=2}^\infty \abs{e_{j} (n)}<\infty$, for $j\in\{a,b\}$. Indeed, our results are 
logarithmically better than Nevai's in the leading term. If $\f{(-1)^n}{n}$ in \eqref{1.3} is 
replaced by $\f{(-1)^n}{n^{1/2}}[\log n]^{-\gamma}$, then Nevai's method requires $\gamma >1$, 
while we require only $\gamma > \f12$. 

\item[(3)] What about other oscillatory potentials like $\f{\cos(\eta n)}{n^\gamma}$ for $\eta\in 
(0,2\pi)$? \eqref{1.3} is the case $\eta =\pi$. Although it is possible his methods extend to this 
case, the conditions in Nevai's paper require cancellations in $b_n + b_{n+1}$ and do not work for 
$\eta\neq\pi$. We will accommodate general $\eta$. 

\item[(4)] Nevai's work suggests that $\f{(-1)^n}{n}$ is akin to $n^{-2}$ potentials, which suggests 
that for $\abs{\alpha}+\abs{\beta}$ small, \eqref{1.3} has finitely many eigenvalues outside $[-2,2]$ 
while for $\abs{\alpha}+\abs{\beta}$ large, it has infinitely many. We will prove the finiteness 
result below. We note that while he does not discuss this case explicitly, Chihara's conditions in 
\cite{Chi} imply finitely many bound states if $a_n=1$ and $\abs{\beta}$ is small. 
\end{SL}

\smallskip
For Jacobi matrices, our main results are:

\begin{t1} Suppose
\begin{align} 
a_n &= 1 + c_n + d_{n+1} -d_n \lb{1.3a}  \\
b_n &= e_n + f_{n+1} - f_n \lb{1.3b} 
\end{align}
with 
\begin{equation} \lb{1.4} 
\sum_{n=1}^\infty \, \abs{c_n} + \abs{e_n} + \abs{d_n}^2 + \abs{f_n}^2 < \infty 
\end{equation}
Then $Z(J)<\infty$ and 
\begin{equation} \lb{1.5} 
\sum_{j,\pm} \sqrt{E_j^\pm (J)^2 -4} < \infty
\end{equation}
where $E_j^\pm (J)$ are the eigenvalues of $J$ in $\pm (2,\infty)$. 
\end{t1}

{\it Remarks.} 1. In case $a_n -1 =\f{\alpha(-1)^n}{n^\gamma} + e_a (n)$, $b_n=\f{\beta(-1)^n}
{n^\gamma} + e_b(n)$, we define
\begin{equation} \lb{1.6} 
d_n =-\sum_{j=n}^\infty \f{\alpha (-1)^n}{n^\gamma} \qquad f_n = -\sum_{j=n}^\infty 
\f{\beta (-1)^n}{n^\gamma}
\end{equation}
Since the sums are $O(n^{-\gamma})$, \eqref{1.4} is then true if $\sum \abs{e_a(n)} + \abs{e_b(n)}<
\infty$ and $\gamma >\f12$. 

\smallskip 
2. If $b_n$ is instead $\f{\cos( \eta n)}{n}$, it is still true that $f_n \equiv -\sum_{j=n}^\infty 
\f{\cos( \eta n)}{n}$ is $O(n^{-1})$, and so in $\ell^2$, and thus this theorem also includes cases 
like $\f{\cos( \eta n)}{n}$ where $b_n + b_{n+1}$ does not have cancellations. 

\smallskip
3. By mimicking the construction of Wigner and von Neumann (see, e.g., 
\cite[Example 1 in Chapter XIII.13]{RS4}, one can construct Jacobi matrices $J$ with 
$a_n\sim 1+ \f{(-1)^n}{n}$ and $b_n\sim \f{(-1)^n}{n}$ as $n\to \infty$ which have $0$ as an 
eigenvalue embedded in the essential spectrum. 

As a converse to Theorem 1, we note 
\begin{t2} Suppose 
\begin{SL}
\item[{\rm{(i)}}] $\limsup [-\sum_{j=1}^n \log (a_j)] >-\infty$
\item[{\rm{(ii)}}] $\sum_{n=1}^\infty (a_n -1)^2 + b_n^2 =\infty$ 
\end{SL}
Then $Z(J)=\infty$. 
\end{t2}

{\it Remark.} If $a_n =1$ (or $a_n=\exp (\f{\alpha (-1)^n}{n^\gamma})$), $b_n =\f{\beta(-1)^n}
{n^\gamma}$, and $\gamma \leq \f12$, then this implies $Z(J)=\infty$, showing $\gamma=\f12$ is the 
borderline. 

\begin{proof} Suppose $Z(J)<\infty$ and (i) holds. Then, by Theorem~1 of Simon-Zlato\v{s} \cite{SZ}, 
\eqref{1.5} holds. A fortiori, the quasi-Szeg\H{o} condition, (1.8) of Killip-Simon \cite{KS}, and 
the $\f32$ Lieb-Thirring bound hold. So, by Theorem~1 of Killip-Simon \cite{KS}, (ii) fails. Thus, 
(i) $+$ $Z(J)<\infty \Rightarrow $ not (ii). So (i) $+$ (ii) $\Rightarrow Z(J) =\infty$. 
\end{proof}

\begin{t3} Suppose $a_n=1$ and \eqref{1.3b} holds with 
\begin{equation} \lb{1.5b} 
\limsup_n \, n^2 [\abs{e_n} + \abs{f_n}^2 + \abs{f_{n+1}}^2 ] <\tfrac18
\end{equation}
Then $J$ has only finitely many bound states. If \eqref{1.3a}, \eqref{1.3b} hold and 
\begin{align} \lb{1.5c} 
\limsup_n \, n^2 \big[ & \abs{c_n} + \abs{c_{n-1}} + 24 \abs{d_{n-1}}^2 + 48 \abs{d_n}^2 + 
24 \abs{d_{n+1}}^2 \notag \\
& + \abs{e_n} + 6 \abs{f_n}^2 + 6 \abs{f_{n+1}}^2 \big] < \tfrac18
\end{align}
then $J$ has finitely many bound states. 
\end{t3}

{\it Remark.} In particular, if $a_n=1$, $b_n=\f{\beta(-1)^n}{n}$, and $\abs{\beta} <\f12$, 
then $J$ has only finitely many bound states. If $\abs{\beta} >1$, it is proven in 
\cite{DHKS} that $J$ has infinitely many bound states. 
Also, if $a_n= 1+\f{\alpha(-1)^n}{n}$ and $b_n$ as before, then for $\beta^2+8\alpha^2< \f{1}{24}$, 
$J$ has also only finitely many bound states, but this bound seems to be far from optimal.

\smallskip
The techniques we will use are two-fold: First, we will use the result of Simon-Zlato\v{s} \cite{SZ} 
that if $-\sum_{j=1}^\infty \log (a_j)$ is conditionally convergent, then \eqref{1.2} holds if and 
only if \eqref{1.5} holds (by a Case-type sum rule). This means that all the results on finiteness 
on $Z(J)$ which we are discussing are equivalent to suitable bounds on eigenvalues. Second, 
to bound eigenvalues, we will use ideas developed in the 1970's to discuss Schr\"odinger operators 
with oscillatory potentials \cite{BC,Cha,CM,Com,CG,Ism,IM,MS,Sar,Sch,Skr}. 
Interestingly, the focus of that work was to handle wild, pathological cases like $V(r)=(1+r)^{-2} 
e^{1/r} \sin (e^{1/r})$ or $V(r)=(1+r)^{-2} e^r \sin (e^r)$, which are extremely unbounded near 
$r=0$ or $r=\infty$ but whose oscillators cause $-\Delta +V$\!, defined by quadratic form methods, to be 
well behaved. In fact, we believe that the most interesting examples are ones like $r^{-1} \sin (r)$ 
which are not unbounded at all, but oscillatory and slowly decaying. 

Most of the 1970's papers discuss scattering or selfadjointness results, although Combescure-Ginibre 
\cite{CG} and Chadan-Martin \cite{CM} do discuss bounds on the number of bound states. Since they 
were not as efficient in using operator bounds, we begin in Section~\ref{s2} with the continuum 
Schr\"odinger operator case.  In Section~\ref{s3}, we discuss the growth of $N(\lambda V)$ as 
$\lambda\to\infty$ for long-range oscillatory potentials. We will prove 

\begin{t4} Let $V_\beta(x) =(1+\abs{x})^{-\beta} \sin (\abs{x})$ for $2>\beta >1$. On $\bbR^\nu$, we have 
\[
-a_- + b_- \lambda^{\nu/\beta} \leq N(\lambda V) \leq a_+ + b_+ \lambda^{\nu/\beta}
\]
for suitable {\rm{(}}$\beta$-dependent{\rm{)}} $a_\pm, b_\pm >0$. 
\end{t4}

We note that if $\beta >2$ so $V_\beta\in L^{\nu/2}$, then it is known (see 
\cite[Theorem~XIII.80]{RS4}) that 
\[
\lim_{\lambda\to\infty} \, \lambda^{-\nu/2} N(\lambda V) = \f{\tau_\nu}{(2\pi)^\nu} 
\int_{V_\beta(x) \leq 0} (-V_\beta (x))^{\nu/2}\, d^\nu x
\]
with $\tau_\nu$ the volume of the unit ball in $\bbR^\nu$. 

\smallskip
In Section~\ref{s4}, we discuss the discrete Schr\"odinger case, that is, Jacobi matrices with 
$a_n \equiv 1$. In Section~\ref{s5}, we discuss the general Jacobi case. The appendix contains 
bounds on the $O(n^{-2})$ situation that we will need in the body of the paper. Since these have not 
been proven in the Jacobi case with optimal constants, it was necessary to include this appendix. 
In particular, in Theorems \ref{TA.5} and \ref{TA.6}, we study Jacobi matrices $J$ with 
$\abs{a_n-1}\sim \f{\gamma_a}{n^2}$ and $\abs{b_n}\sim \f{\gamma_b}{n^2}$ and discuss finiteness 
(resp.\ infinitude) of the discrete spectrum of $J$ in $[-2,2]^\mathrm{c}$, depending on whether 
$2\gamma_a +\gamma_b< \f14$, (resp.\ $2\gamma_a +\gamma_b > \f14)$, thereby extending results of 
Chihara in \cite{Chi}. 

\smallskip
We would like to thank Rowan Killip, Paul Nevai, Mihai Stoiciu, and Andrej Zlato\v{s} for 
valuable communications.

\medskip

\section{The Continuum Schr\"odinger Case} \lb{s2} 

Let $W$ be an $\bbR^\nu$-valued $C^1$ function on $\bbR^\nu$ or a piecewise $C^1$ continuous function 
on $\bbR$ so that $\nabla \cdot W$ is also bounded. In fact, once one has the bounds below, it is 
easy to accommodate arbitrary distributions $W$ with $W\in L^\nu +L^\infty$ (when $\nu\geq 3$) even 
if $\nabla \cdot W$ is not bounded. For our applications of interest, we make these simplifying 
assumptions. 

\begin{proposition}[Combescure-Ginibre Lemma \cite{CG}]\lb{P2.1} If $\varphi\in C_0^\infty$, 
\begin{equation} \lb{2.1} 
\abs{\langle \varphi, \nabla W\varphi\rangle} \leq 2 \|W\varphi\|\, \|\nabla \varphi\|
\end{equation}
\end{proposition}

\begin{proof} First, integrate by parts, $\langle\varphi, \nabla W \varphi\rangle =2 \Real \langle 
W\varphi, \nabla\varphi\rangle$. Then use the Schwarz inequality. 
\end{proof} 

\begin{theorem} \lb{T2.2} Let $H=-\Delta + V_1 + \nabla \cdot W$ and $H_1 =-\Delta + 2V_1 - 4W^2$. 
Then 
\begin{equation} \lb{2.2} 
H \geq \tfrac 12\, H_1
\end{equation}
In particular, if $N(V)$ is the number of negative eigenvalues of $-\Delta + V$\!, then 
\begin{equation} \lb{2.3} 
N(V_1 + \nabla \cdot W) \leq N (2V_1 - 4W^2)
\end{equation}
and if $E_p (V) = \tr([-HE_{(-\infty, 0]}(H)]^p)$ for $H=-\Delta + V$ is the Lieb-Thirring sum 
of eigenvalue powers, then 
\begin{equation} \lb{2.4}
E_p (V_1 + \nabla\cdot W)\leq 2^{-p} E_p (2V_1 - 4W^2) 
\end{equation}
\end{theorem}

\begin{proof} By \eqref{2.1}, 
\begin{align*} 
\langle\varphi, (-\Delta + V_1 + \nabla W)\varphi\rangle &\geq \langle \varphi, (-\Delta + V_1) 
\varphi\rangle - \veps \langle\varphi, -\Delta\varphi\rangle  - \veps^{-1} \langle \varphi, W^2 \varphi 
\rangle \\ 
&= \left\langle \varphi, (1-\veps)\biggl(-\Delta +  \f{1}{1-\veps} \, V_1 - \f{1}{\veps (1-\veps)} \, 
W^2\biggr) \varphi \right\rangle 
\end{align*} 
In the absence of a $V_1$ term, the optimal choice of $\veps$ is $\veps =\f12$ (to minimize $\f{1}
{\veps(1-\veps)}$), so we make that choice in general. It yields \eqref{2.2}, which in turn immediately 
implies \eqref{2.3} and \eqref{2.4}. 
\end{proof} 

The bound \eqref{2.1} and its proof are taken from Combescure-Ginibre \cite{CG}. While they use 
the Schwarz inequality, they do not explicitly note \eqref{2.2}, which causes them to make extra 
arguments that can be less efficient than using \eqref{2.2}. For example, if $V_1 =0$, $\nu =3$,  
and 
\[
w^2 = (4\pi)^{-2} \int d^3 x\, dy^3 \, \f{W^2(x) W^2 (y)}{\abs{x-y}^2}
\]
then \eqref{2.3} and the Birman-Schwinger principle immediately imply that 
\[
N(\nabla \cdot W) \leq 16 w^2
\]
while Combescure-Ginibre \cite{CG} only claim 
\[
N(\nabla \cdot W) \leq 16 w^2 (1+w)^2
\]
which is much worse for large $w$. 

For $\nu =1$, $V_1 =0$, \eqref{2.3} is a result of Chadan-Martin \cite{CM} who use Sturm comparison 
methods rather than the Schwarz inequality and the Combescure-Ginibre lemma.  Theorem~\ref{T2.2} has 
some immediate consequences:

\begin{corollary} \lb{C2.3} For $\nu \geq 3$, 
\begin{equation} \lb{2.5} 
N(V_1 + \nabla\cdot W) \leq c_\nu (\|V_1\|_{\nu/2}^{\nu/2} + \|W\|_\nu^\nu)
\end{equation}
For general $\nu$ and $p\geq\f12$ if $\nu =1$, $p>0$ if $\nu =2$, and $p\geq 0$ if $\nu \geq 3$, 
\begin{equation} \lb{2.6} 
E_p (V_1 + \nabla W) \leq c_{\nu,p} (\|V_1\|_{p+\nu/2}^{p + \nu/2} + \|W\|_{2p+\nu}^{2p +\nu})
\end{equation}
\end{corollary} 

\begin{proof} \eqref{2.5} is just the Cwikel-Lieb-Rozenblum \cite{Cw,L1,Roz} bound, given \eqref{2.3}. 
\eqref{2.6} is the Lieb-Thirring bound \cite{LT1,LT2} when $p$ is strictly larger than the minimal 
value. $p=0$ for $\nu\geq 3$ is \eqref{2.5} while $p=\f12$, $\nu=1$ is due to Weidl \cite{We} 
(see also Hundertmark-Lieb-Thomas \cite{HLT}). 
\end{proof} 

If $\nu\geq 3$ and $V\in L^{\nu/2}$, we have $N(\lambda V)\leq c\lambda^{\nu/2}$, but \eqref{2.5} 
only implies that 
\[
N(\lambda (V_1 + \nabla W) ) \leq c_1 \lambda^{\nu/2} + c_2\lambda^\nu
\]
In the next section, we see that in some specific cases, $N(\lambda V)$ really does grow at rates  
arbitrarily close to $\lambda^\nu$. 

\begin{corollary} \lb{C2.4} If $\nu\neq 2$ and $4W^2 -2V_1 \leq \f{(\nu-2)^2}{4} \abs{x}^{-2}$, then 
$-\Delta + V_1 + \nabla W_1$ has no bound states. If $4W^2 < \f{(\nu-2)^2}{4} \abs{x}^{-2}$ and 
$V_1\in L^{\nu/2}$ {\rm{(}}if $\nu\geq 3${\rm{)}} or $\int (1+\abs{x}) \abs{V_1 (x)}\, dx < \infty$ 
{\rm{(}}$\nu=1${\rm{)}}, then $-\Delta + V_1 + \nabla \cdot W_1$ has finitely many bound states. 
\end{corollary} 

\begin{proof} The first statement is immediate from \eqref{2.3} and Theorem~\ref{TA.3}. The second 
follows from 
\[
-\Delta + 2V_1 - 4W^2 = -(1-\veps)\Delta -4W^2 - \veps\Delta + 2V_1
\]
and the Cwikel-Lieb-Rozenblum estimate if $\nu\geq 3$ and Bargmann's bound if $\nu =1$. 
\end{proof}

\begin{example} \lb{E2.5} If $V =\f{\sin(r)}{r^\alpha}$, we can write $V=V_1 + \nabla\cdot W$ 
where $W=\hat r (-\f{\cos(r)}{r^\alpha} - \f{\alpha\sin(r)}{r^{\alpha+1}}) f(r)$ where $f\in C^\infty$ 
vanishes near $0$ and is $1$ near $\infty$. This shows that for $\alpha <1$, $-\Delta + \lambda V_0$ 
has finitely many bound states for all $\lambda$, and when  $\alpha =1$, it has finitely many bound 
states if $\abs{\lambda}$ is small. An argument similar to that in \cite{DHKS} shows that if 
$\alpha =1$ and $\lambda$ is large, $-\Delta + \lambda V$ has an infinity of negative eigenvalues. \qed
\end{example}

\medskip

\section{Schr\"odinger Operators at Large Coupling} \lb{s3}

Our purpose here is to prove Theorem~4 that $V_\beta =\f{\sin(r)}{(1+r)^\beta}$ (with $2>\beta >1$) has 
$N(\lambda V_\beta)$ growing as $\lambda^{\nu/\beta}$ for $\lambda$ large. We give the details when $\nu=1$ 
on a half-line and then discuss the case when $\nu\geq 2$. 

\begin{proof}[Proof of Theorem~4] Half-line case: We begin with the upper bound. Let $\varphi_R(r)$ 
be a $C^\infty$ function with $\|\varphi_R\|_\infty =1$, which is $0$ if $r<R$ and $1$ if $r>R+1$. 
By translation, we may assume the derivatives $\f{d^\alpha\varphi_R}{dx^\alpha}$ are uniformly bounded 
in $x$ and $R$ (for fixed $\alpha$). Let $W_R(r)=-\int_r^\infty \varphi_R(s) V_\beta(s)\, ds$ and $V_{1,R}
= V_\beta -\f{d}{dr} W_R$. Define $V_{2,R}(r)=\max_{s\geq r}\, \abs{V_{1,R}(s)}$. Then 
\begin{align}
-\f{d^2}{dx^2} + \lambda V_\beta &= \biggl( -\f12 \, \f{d^2}{dx^2} + \lambda V_{1,R}\biggr) + 
\biggl( -\f12 \, \f{d^2}{dx^2} + \lambda\, \f{dW_R}{dr} \biggr) \notag \\
&\geq \biggl( -\f12\, \f{d^2}{dx^2} - \lambda V_{2,R}\biggr) + \f12\, \biggl( -\f12 \, \f{d^2}{dx^2} 
- 8\lambda^2 W_R^2 \biggr) \lb{3.1} 
\end{align}

Next, note that 
\begin{equation} \lb{3.2} 
W_R(r) \leq C (\max (r,R))^{-\beta}
\end{equation}
We also have 
\begin{equation} \lb{3.3} 
\abs{V_{2,R}(r)} \leq r^{-\beta}
\end{equation}
and is zero if $r>R +1$. 

Calogero \cite{Cal} has proven that if $V$ is monotone decreasing and nonnegative, then $N(-V)\leq 
2\pi^{-1} \int_0^\infty \abs{V(s)}^{1/2}\, ds$. This bound, \eqref{3.1}--\eqref{3.3}, and the fact 
that $\dim (E_{(-\infty, 0)}(A+B))\leq \dim E_{(-\infty, 0)}(A) + \dim E_{(-\infty, 0)}(B)$ (by the 
variational principle) imply that for any $R$, 
\begin{align*} 
N(\lambda V_\beta) &\leq C_1 \biggl[ \lambda^{1/2} \int_0^{R+1} r^{-\beta/2}\, dr + \lambda \int_0^\infty 
\max(R, r)^{-\beta}\, dr \biggr] \\
&= C_2 [\lambda^{1/2} R^{1-\beta/2} + \lambda R^{1-\beta}]
\end{align*}
since $1 < \beta < 2$. Pick $R=\lambda^{1/\beta}$ and get 
\[
N(\lambda V_\beta) \leq 2C_2 \lambda^{1/\beta}
\]

On the other side, consider the operator $\tilde H(\lambda)$, which is $-\f{d^2}{dx^2} + \lambda 
V_\beta$ with Dirichlet boundary conditions added at the points $(2n+\f32)\pi\pm\f{\pi}{3}$. Adding 
such boundary conditions only increases the operator, so $N(\lambda V_\beta)\geq \#$ of negative 
eigenvalues of $\tilde H(\lambda)$. In each interval of the form $[(\f{2n+3}{2})\pi - \f{\pi}{3}, 
(\f{2n+3}{2}) \pi + \f{\pi}{3}]$, $\sin(r)$ is less than $-\f12$, so $V_\beta \leq -\f{\lambda}
{2[(2n+3)\pi]^\beta}$ on the entire interval. The lowest Dirichlet eigenvalue of $-\f{d^2}{dx^2}$ on 
such an interval is $\f94$, so each interval with 
\[
\f{\lambda}{2[(2n+3)\pi]^\beta} > \f94
\]
contributes an eigenvalue so 
\[
N(\lambda V_\beta) \geq C_3 \lambda^{1/\beta}
\]
This completes the proof of Theorem~4 in the half-line case. 

\smallskip
One might think that it would help to use the fact that small $n$ intervals provide 
$O(\lambda^{1/2})$ eigenvalues rather than just the $1$ we use, but a detailed analysis shows it 
improves the constant in front of $\lambda^{1/\beta}$ but not the power. 

\smallskip
Higher dimensions: The lower bound is similar to the half-line case. We have 
$\sin(r)< -\f12$ on annuli which we can partially cover with suitable disjoint cubes 
of fixed size, finding cubes where $V$ is deep enough when the distance of the cube 
from the origin is no more than $C\lambda^{1/\beta}$. The number of such cubes is 
$O(\lambda^{\nu/\beta})$ so we get an $O(\lambda^{\nu/\beta})$ lower bound. 

For the upper bound when $\nu\geq 3$, we can replace Calogero's bound with the Cwikel-Lieb-Rozenblum 
bound. Since $\int_0^R r^{-\beta\nu/2}\, d^\nu r =C_4 R^{\nu(1-\beta/2)}$ and $\int_R^\infty 
r^{-\beta\nu}\, d^\nu r =C_5 R^{\nu(1-\beta)}$, we find 
\[
N(\lambda V_\beta) \leq C_6 [\lambda^{\nu/2}R^{\nu(1-\beta/2)} + \lambda^\nu R^{\nu(1-\beta)}]
\]
so picking $R=\lambda^{1/\beta}$, we get $N(\lambda V_\beta)\leq C_7 \lambda^{\nu/\beta}$. 

$\nu=2$ is messier. We will sketch the idea, but omit the details. One needs to use the spherical 
symmetry and consider each partial wave separately. By using the analog of \eqref{3.1}, we see, on 
functions of angular momentum $\ell$, there is an effective potential which bounds $-\Delta + \lambda 
V_\beta$ from below, viz, 
\[
V_{\ell,\eff}=-\f{1}{4r^2} + \f{\ell^2}{4r^2} - \lambda r^{-\beta} \chi_{(0,R+1)} - 
\lambda^2 \max (r, R)^{-2\beta}
\]
We need to consider three regions: 
\begin{SL} 
\item[(i)] $\ell \geq C_8 \lambda^{1/\beta}$: Take $R=\lambda^{1/\beta}$ and find $V_{\ell, \eff} \geq 
0$ so there are no bound states. 
\item[(ii)] $1\leq \ell\leq C_8 \lambda^{1/\beta}$: We take $R=\lambda^{1/\beta}$, drop the  
$\f{\ell^2 -4}{4r^2}$ term, and use Calogero's bound to get a bound per partial wave of $C_9 
\lambda^{1/\beta}$ as in the one-dimensional case. 
\item[(iii)] $\ell=0$: The singularity of $-r^{-2}$ at both $0$ and infinity requires us to place 
Dirichlet boundary conditions at $1$ and a point $R_2 =\lambda^{2/\beta-1}$, which for large $\lambda$ 
is much larger than $R_1 =\lambda^{1/\beta}$ (since $\f{1}{\beta} < 1 < \f{1}{\beta-1} < 
\f{2}{\beta-1}$). On $(R_2, \infty)$, we can use the fact that $\lambda V_\beta\geq -\f14 \, \f{1}
{r^2 \log r}$ and Theorem~\ref{TA.2} to see the Dirichlet operator has no bound states. On $(0,1)$, 
we can bound the $\ell=0$ states by all states for the Dirichlet Laplacian in $L^2 (\{\abs{x} <1\}, 
d^2 x)$ with Dirichlet boundary conditions with energy below $c_a \lambda$ (where $c= \max_{\abs{r}
\leq 1} -\f{\sin(r)}{(1+r)^\beta}$). It is known (by Weyl's theorem, see \cite[p.~271]{RS4}) that this 
is asymptotically $c_{10}\lambda$ since $\nu=2$. In $(1,R_2)$, we can use Calogero's bound where now 
``$V$" is $-\f{1}{4r^2}- \lambda V_{2,R} - 4\lambda^2 W_R^2$. We get a bound by $c_{11} \int_1^{R_2} 
\f{dr}{r} + c_{12} \lambda^{1/\beta}$. Taking into account the possible two states lost by adding the 
Dirichlet boundary conditions in the $\ell=0$ space, we get 
\[
N(\lambda V_\beta) \leq c_{13}  \lambda^{1/\beta}(\lambda^{1/\beta}) + c_{14} (\lambda + \lambda^{1/\beta} 
+ \log (\abs{\lambda}+1)) 
\]
which is the required large $\lambda^{2/\beta}$ bound.  
\end{SL}
\end{proof}
\medskip

\section{Discrete Schr\"odinger Operators} \lb{s4}

Our main goal in this section is to extend \eqref{2.1} and Theorem~\ref{T2.2} to the discrete case. 
It will be convenient to consider operators on all of $\bbZ$ and get bounds on Jacobi matrices by 
restriction. We will also restrict to eigenvalues above energy $2$. One can then control energies 
below $-2$ by using 
\begin{equation} \lb{4.1} 
U_0 J(\{a_n\}, \{b_n\}) U_0^{-1} = -J (\{a_n\}, \{-b_n\})
\end{equation}
where $J(\{a_n\}, \{b_n\})$ is the Jacobi matrix \eqref{1.1} with parameters $a_n$, $b_n$, and 
\begin{equation} \lb{4.1a} 
(U_0u)(n) = (-1)^n u(n)
\end{equation}

On $\ell^2 (\bbZ)$, define two operators $H_0$ and $\delta_+$ as 
\begin{align} 
(H_0u)(n) &= u(n+1) + u(n-1) \lb{4.2}  \\
(\delta_+ u)(n) &= u(n+1) - u(n) \lb{4.3} 
\end{align} 
Then $\delta_- \equiv \delta_+^*$ is given by 
\[
(\delta_-u)(n) = u(n-1) - u(n) 
\]
and 
\begin{equation} \lb{4.4} 
\delta_+ \delta_- = \delta_- \delta_+ = 2-H_0
\end{equation}
(for if $\delta_+ =R-1$ and $\delta_- =L-1$, then $RL=LR=1$ and $H_0 =L+R$). Let $b_n$ and $f_n$ be 
sequences on $\bbZ$ and suppose 
\begin{equation} \lb{4.5} 
b_n =f_{n+1} - f_n = (\delta_+f)_n 
\end{equation}
Then in $\ell^2 (\bbZ)$, for $u$ real and of finite support, 
\begin{align}
\langle u,bu\rangle &= \sum_n b_n \abs{u(n)}^2 \notag \\
&= \sum_n (f_{n+1}-f_n) \abs{u(n)}^2 \notag \\
&= \sum_n f_n (\abs{u(n-1)}^2 - \abs{u(n)}^2) \notag \\
&= \langle \delta_- u, f(1+L)u\rangle \lb{4.6} 
\end{align}
Since $\|\delta_- u\|^2 = \langle u, \delta_+ \delta_- u\rangle = \langle u, (2-H_0)u\rangle$ by 
\eqref{4.4}, we see that 

\begin{lemma} \lb{L4.1} If $b$ is given by \eqref{4.5}, then 
\begin{equation} \lb{4.7} 
\abs{\langle u, bu\rangle} \leq \langle u, (2-H_0)u\rangle^{1/2} [2 \langle u, (f^2 + \tilde f^2) 
u\rangle ]^{1/2}
\end{equation}
where 
\begin{equation} \lb{4.8} 
\tilde f_n = f_{n+1}
\end{equation}
\end{lemma} 

\begin{proof} In getting \eqref{4.7}, we used \eqref{4.4}, \eqref{4.6}, 
\[
\|f(1+L)u\|^2 \leq 2\|fu\|^2 + 2\|fLu\|^2 = 2\|fu\|^2 + 2\|\tilde f u\|^2
\]
and the fact that, because of $\abs{\abs{x}-\abs{y}}\le \abs{x-y}$, we also have 
\[
\langle \abs{u}, (2-H_0)\abs{u}\rangle = \tfrac12 \sum_n \abs{\abs{u(n+1)}-\abs{u(n)}}^2 
\le 
\langle u, (2-H_0)u\rangle 
\]
so it suffices to prove the result for real valued sequences $u$. 
\end{proof}

We will later need the following estimate that was proven along the way (we get $J_0$ by restricting to 
$u$'s of support on $\bbZ_+$): 
\begin{align} 
\abs{\langle u, \delta_+f\, u\rangle} &\leq 2 \abs{\langle u, (2-H_0)u\rangle}^{1/2} 
\abs{\langle u, \tfrac12\, (f^2 + \tilde f^2)u\rangle}^{1/2} \lb{4.8a}  \\
&\leq \veps \langle u, (2-J_0)u\rangle + \veps^{-1} \langle u, \tfrac12\, (f^2 + \tilde f^2) 
u\rangle \lb{4.8b}
\end{align}

\begin{theorem} \lb{T4.2} Let $b_n$ be a sequence on $\bbZ_+$ so that $\lim_{n\to\infty} \sum_{j=1}^n 
b_j$ exists, and let
\begin{equation} \lb{4.9} 
f_n =-\sum_{j=n}^\infty b_j
\end{equation}
Let $J$ be the Jacobi matrix with $a_n \equiv 1$ and $b$'s given by $b_n$. Let $J^\pm$ be the Jacobi 
matrix with $a_n \equiv 1$ and $b$'s given by 
\begin{equation} \lb{4.9a} 
\pm 2 (f^2 + \tilde f^2)
\end{equation}
Then
\begin{SL} 
\item[{\rm{(i)}}] 
\begin{equation} \lb{4.10} 
\dim E_{(2,\infty)} (J) \leq \dim E_{(2,\infty)} (J^+)
\end{equation}
\item[{\rm{(ii)}}] 
\begin{equation} \lb{4.11} 
\dim E_{(-\infty, -2)} (J) \leq \dim E_{(-\infty, -2)}(J^-)
\end{equation}
\item[{\rm{(iii)}}] If the eigenvalues $E$ of $J^\pm$ outside $[-2,2]$ obey 
\begin{equation} \lb{4.12} 
\sum_j (\abs{E_j (J^\pm)}-2)^\alpha <\infty
\end{equation}
for some $\alpha$ and for both $J^+$ and $J^-$, then 
\begin{equation} \lb{4.13} 
\sum_j (\abs{E_j(J)}-2)^\alpha < \infty
\end{equation}
\end{SL} 
\end{theorem}

\begin{proof} Define 
\[
b_0 =-\sum_{j=1}^\infty b_j 
\]
so if $f$ is extended to $\bbZ$ by setting $f_k =0$ for $k\leq 0$, we have $b=\tilde f -f$. Thus, 
by \eqref{4.7} as operator on $\ell^2 (\bbZ)$, 
\[
-b \geq\tfrac12\, [-(2-H_0) -2 (f^2 + \tilde f^2)]
\]
so 
\[
2-H_0 -b \geq \tfrac12\, [(2-H_0)-2 (f^2 + \tilde f^2)]
\]

Now restrict to functions supported on $\bbZ_+$ to get 
\begin{equation} \lb{4.14} 
2 - J \geq \tfrac12\, [2-J^+]
\end{equation}
This yields \eqref{4.10} and \eqref{4.1} then yields \eqref{4.11}. The two together imply \eqref{4.13}. 
\end{proof} 

\smallskip
\noindent{\bf Example.} Let $b_n =\f{\beta(-1)^n}{n}$. Then $f_n \sim -\f12\, \f{\beta (-1)^n}{n} + 
O(\f{1}{n^2})$ and the leading term in $2n^2(f^2 + \tilde f^2) = 4(\f12 \beta)^2 = \beta^2$. By 
Theorem~\ref{TA.5}, if $\beta^2 < \f14$, $J(a=1,b)$ has finitely many eigenvalues, that is, 
$\abs{\beta} <\f12$ produces finitely many eigenvalues. 

On the other hand, if $\abs{\beta} >1$, it is known \cite{DHKS} that $H$ has an infinite 
number of  bound states. It would be interesting to determine the exact value of the 
coupling constant, where the shift from finitely many to infinitely many bound states 
takes place.

\begin{proof}[Proof of the First Assertion in Theorem~3] By \eqref{4.14}, if $b_n$ has the form 
\eqref{1.3a} and $J^\pm$ is formed with $b_n^\pm = 2e_n^\pm \pm 2(f_n^2 + f_{n+1}^2)$, then 
\[ 
2 \mp J \geq \tfrac12\, [2\mp J^\pm ]
\]
If \eqref{1.5b} holds, then 
\[ 
\limsup_n \, n^2 [ \abs{\tilde b_n^\pm} ] < \tfrac14
\]
so $J^\pm$ have finitely many bound states by Theorem~\ref{TA.5}. 
\end{proof}

\medskip

\section{Oscillatory Jacobi Matrices} \lb{s5} 

In this section, we will prove Theorems~1, 2, and 3 by accommodating general values of $a_n$ 
within the bounds of the last section. Recall that $R$ acts as $Ru(n)= u(n+1)$ and we defined 
$\delta_+ =R-1$. It will be convenient to write the Jacobi matrix $J=aR+ R^*a +b$ in ``divergence 
form,'' that is, write it as $J= -\delta_+^* g\delta_+ +q$. Let us first consider the whole-line 
case: Given sequences $a,b$ on $\bbZ_+$, we extend them to sequences on $\bbZ$ by setting 
$a_n = 1$ and $b_n = 0$ for $n \le 0$. 
We denote the corresponding operator on $\ell^2(\bbZ)$ by $K$. With $a^\sharp= R^* aR$, that is, 
$a^\sharp_n= a_{n-1}$, and $-\delta_+^* g\delta_+ = -R^*gR +gR +R^* g -g $, we see that $g=a$ 
and $q= b+a+a^\sharp$. Thus, recalling $\delta_+^*\delta_+= 2-H_0$, 
\begin{align*}
K 
&= 
-\delta_+^*a\delta_+ + b+ a+a^\sharp \\
&= H_0 + b+ a+a^\sharp -2 - \delta_+^* (a-1)\delta_+ 
\end{align*}
which shows 
$$ 
\langle u,Ku \rangle 
= 
\langle u, H_0 u\rangle + \langle u, (b+ a+ a^\sharp -2)u \rangle 
- \langle \delta_+ u, (a-1)\delta_+ u\rangle
$$
By restriction to $u$'s supported on $\bbZ_+$, we get
\begin{equation} \lb{5.1} 
\langle u,Ju \rangle 
= 
\langle u, J_0 u\rangle + \langle u, (b+ a+ a^\sharp -2)u \rangle 
- \langle \delta_+ u, (a-1)\delta_+ u\rangle
\end{equation}
where one should keep in mind that $a^\sharp_1 = 1$.

We first estimate the third term in \eqref{5.1}. 
Writing $a$ as in (\ref{1.3a}), that is, $a= 1+c +\delta_+ d$, it reads 
\begin{equation}\lb{5.1b}
\langle \delta_+ u, (a-1)\delta_+ u\rangle 
=
\langle \delta_+ u, c\delta_+ u\rangle + 
\langle \delta_+ u, (\delta_+ d)\,\delta_+ u\rangle 
\end{equation}
With $(x)_-= \max(-x,0)$, the negative part, we have 
\begin{align}
\langle \delta_+ u, c\delta_+ u\rangle 
&\ge 
- \langle \delta_+ u, c_- \delta_+ u\rangle \nonumber \\
&= 
-\sum_{n} (c_n)_- \big( \abs{u(n+1)}^2 +\abs{u(n)}^2 -2\Real(\ol{u(n+1)}u(n)) \big)
\nonumber \\
&\ge - \sum_{n} 2(c_n)_- \big( \abs{u(n+1)}^2 +\abs{u(n)}^2 \big) 
\nonumber \\
&= -2\langle u, (c_-+ c_-^\sharp) u\rangle \label{5.1c}
\end{align}
where one should keep in mind that $c^\sharp_1= 0$. 
For the last term in \eqref{5.1b}, we note that by \eqref{4.8b}, 
\begin{displaymath}
\langle \delta_+ u,  (\delta_+ d) \delta_+ u \rangle  
\ge 
- \langle \delta_+ u, A\delta_+ u\rangle 
\end{displaymath}
where
\begin{displaymath}
A = A_\veps = \veps (2-J_0) + \frac{1}{2\veps}\, (d^2 + \tilde d^2)
\end{displaymath}
Now since $A\geq 0$, 
\begin{align*}
\langle \delta^+ u, A\delta^+u\rangle &= \|A^{1/2} (R-1)u\|^2 \notag \\
&\leq 2\|A^{1/2} Ru\|^2 + 2\|A^{1/2} u\|^2 \notag \\
&= \langle u, [2(R^* AR) + 2A]u\rangle 
\end{align*}
We have $R^* J_0 R = J_0$, $R^* \tilde f R = f$, and $R^*f R = f^\sharp$. 
Thus we arrive at 
\begin{equation} \lb{5.6} 
\langle \delta_+ u,  (\delta_+ d) \delta_+ u \rangle  
\ge 
- \langle u,Bu\rangle
\end{equation}
with 
\begin{displaymath}
B=4\veps (2-J_0) + \veps^{-1} [(d^\sharp)^2 + 2d^2 + \tilde d^2 ]
\end{displaymath}
Writing $b= e+\delta_+ f$, as in \eqref{1.3b}, and putting \eqref{5.1}, \eqref{5.1b}, \eqref{5.1c}, 
and \eqref{5.6} together, we have 
\begin{align} \label{5.7b} 
\langle u, (2-J)u\rangle \ge 
& 
(1-4\veps)\langle u, (2-J_0)u\rangle \nonumber\\
&- \langle u,(e+\abs{c}+\abs{c^\sharp} + \veps^{-1}((d^\sharp)^2+2d^2+\tilde{d}^2)) u \rangle 
\nonumber \\
&
- \langle u, (\delta_+f + (\delta_+d)^\sharp +\delta_+d )u\rangle
\end{align}
Estimating the last term in \eqref{5.7b} again with the help of \eqref{4.8b} yields 
\begin{align} \label{5.9a} 
\langle u, (2-J)u\rangle \ge 
& 
(1-(\mu+\nu+4\veps))\langle u, (2-J_0)u\rangle \nonumber\\
& 
- \langle u, (e+\abs{c}+\abs{c^\sharp})u \rangle 
- \frac{1}{2\nu} \langle u, (f^2 +\tilde{f}^2)u \rangle \nonumber \\
& 
- \biggl(\frac{1}{\mu}+\frac{1}{\veps}\biggr) \langle u, ((d^\sharp)^2 +2d^2 +\tilde{d}^2)u \rangle 
\end{align} 

Choosing $\mu=\nu=\veps= \tfrac{1}{12}$, we get 
\begin{equation} \lb{5.11} 
2-J \geq \tfrac12\, [2-J_0 -W] 
\end{equation}
where 
\begin{equation} \lb{5.12} 
W = 2e+ 2\abs{c} + 2\abs{c^\sharp} + 12 [(f)^2 + (\tilde f)^2] 
    + 48 [(d^\sharp)^2 + 2(d)^2 + (\tilde d)^2 ] 
\end{equation}

\begin{proof}[Proof of Theorem~1] \eqref{1.4} and \eqref{5.12} imply 
\[
\sum\, \abs{W_n} < \infty
\]
Thus, by Hundertmark-Simon \cite{HS}, \eqref{1.5} holds for the eigenvalues of $J_0\pm W_0$. 
By \eqref{5.11}, \eqref{4.1}, and the min-max principle, \eqref{1.5} holds for $J$. 

Moreover, by \eqref{1.3} and \eqref{1.4}, $\sum (a_n-1)$ is conditionally convergent and, 
by \eqref{1.4}, $\sum (a_n -1)^2 <\infty$. It follows that $\sum\log (a_n)$ is conditionally 
convergent. Thus, by Theorem~1 of Simon-Zlato\v{s} \cite{SZ}, $Z(J)<\infty$. 
\end{proof} 

\begin{proof}[Proof of Theorem~3] By \eqref{5.12} and \eqref{1.5c}, for large $n$, 
\[
\abs{W_n} \leq \f{1-\veps}{4n^2} 
\]
for some $\veps >0$. It follows, by Theorem~\ref{TA.5}, that $J_0\pm W$ has only finitely many 
bound states. Hence, by \eqref{4.1}, \eqref{5.11}, and the min-max principle, $J$ has finitely 
many bound states. 
\end{proof}

\medskip
\appendix
\section{Finiteness of the Eigenvalue Spectrum for Potentials 
of a Definite Sign} \lb{App}
\renewcommand{\theequation}{A.\arabic{equation}}
\renewcommand{\thetheorem}{A.\arabic{theorem}}
\setcounter{theorem}{0}
\setcounter{equation}{0}

We need information on finiteness results for nonoscillatory potentials. For Schr\"odinger operators, 
these results are well known, but we include some discussion here for two reasons: Optimal constants 
for Jacobi matrices are not known. The weak $L^{\nu/2}$ results we discuss are new. We begin with 
a version of Hardy's inequality with optimal constant: 

\begin{theorem} \lb{TA.1} Let $H_0 =-\f{d^2}{dx^2}$ on $L^2 (0,\infty)$ with $u(0)=0$ boundary 
conditions. Let $V$ be a bounded function on $[0,\infty)$ with $V(x)\to 0$ at infinity. Then 
\begin{SL} 
\item[{\rm{(i)}}] If $V(x) \geq (4x^2)^{-1}$ for all {\rm{(}}resp.~all large{\rm{)}} $x$, then 
$H_0 +V$ has no {\rm{(}}resp.~finitely many{\rm{)}} bound states. In particular, for any $\varphi 
\in Q(H_0)$, the form domain of $H_0$, 
\begin{equation} \lb{A.1} 
\int \f{\abs{\varphi(x)}^2}{4x^2} \, dx \leq \int \abs{\nabla\varphi(x)}^2\, dx 
\end{equation}
This is known as Hardy's inequality. 

\item[{\rm{(ii)}}] If $V(x) \leq -(1+\veps) (4x^2)^{-1}$ for $x>R_0$ and some $R_0,\veps >0$, 
then $H_0 +V$ has an infinite number of bound states. 
\end{SL} 
\end{theorem} 

{\it Remark.} We only assume $V$ bounded to avoid technicalities. In fact, one can use \eqref{A.1} 
to discuss $V$'s with $V(x) \geq -\f14 x^{-2} -c$. 

\begin{proof} Sturm's theory (see \cite[pp.~90--94]{RS4}) says that the number of negative eigenvalues 
of $H_0 +V$ is precisely the number of zeros of $-u'' (x) + V(x) u(x)=0$, $u(0)=0$, and that any 
other solution of $-w'' + Vw =0$ has a zero between any two zeros of $u$, and vice-versa. Thus, if  
some solution, $w$, of $-w''+Vw=0$, is positive, $H_0 +V$ has no eigenvalues, and if it has an infinity 
of zeros, $H_0 +V$ has an infinity of eigenvalues. 

$u(x)=x^{1/2}$ solves $-u'' - \f14 x^{-2} u=0$, showing (i). On the other hand, $u(x)=x^\alpha$ with 
$\alpha (\alpha -1) =\f14 (1+\veps)$ solves $-u'' - \f14 (1+\veps) x^{-2} u=0$. If $\veps >0$, $\alpha$ 
has an imaginary part and $\Real (x^\alpha)$ has an infinity of zeros. This plus a comparison theorem 
implies the results. 
\end{proof}

{\it Remark.} There are two other ways to prove Hardy's inequality: Let $a=\f{d}{dx} -\f{1}{2x}$. 
Then $a^*a =H_0 -(4x^2)^{-1}$; a careful version of this proof requires consideration of boundary 
conditions at $x=0$. 
Second (Herbst \cite{Her}), \eqref{A.1} is equivalent to $x^{-1} p^{-2} x^{-1} \leq 4$. 
Changing variables from $x$ to $e^u =x$, using the explicit form $p^{-2}(x,y)= \max(x,y)$ for 
the kernel of $p^{-2}$ with Dirichlet boundary conditions at zero, one gets that $x^{-1} p^{-2} x^{-1}$ 
is unitarily equivalent to convolution with $e^{-\f12 \abs{u}}$ on $L^2 (\bbR)$. This operator has 
norm $\int_{-\infty}^\infty e^{-\f12 \abs{u}}\, du=4$. This argument also shows that the operator 
has continuous spectrum, so if its norm is larger than $1$, a Birman-Schwinger-type argument provides 
an alternate proof of an infinity of bound states. 

\smallskip
For reasons that will become clear when we discuss the two-dimensional case, we need more on the 
borderline $-\f14 x^{-2}$ case. 

\begin{theorem}\lb{TA.2} Let 
\begin{equation} \lb{A.2} 
X_\gamma (x) = -\frac{1}{4x^2} - \gamma \chi_{(2,\infty)}(x) \, \f{1}{x^2 (\log x)^2}
\end{equation}
Let $V$ be a bounded function on $[0,\infty)$ with $V(x)\to 0$ as $x\to\infty$. Then 
\begin{SL} 
\item[{\rm{(i)}}] If $V(x) \geq X_{\gamma=1/4}(x)$ for large $x$, then $H_0 +V$ has finitely many 
bound states. 
\item[{\rm{(ii)}}] If $V(x) \leq X_\gamma (x)$ for some $\gamma >\f14$ and all large $x$, then $H_0 
+V$ has infinitely many bound states. 
\end{SL} 
\end{theorem}

\begin{proof} Let $u_\alpha =x^{1/2} (\log x)^\alpha$ in the region $x>2$. Then 
\[
-u'' + X_{\gamma =-\alpha (\alpha -1)} u=0
\]
by a direct calculation. The proof is now identical to that of Theorem~\ref{TA.1}. 
\end{proof} 

\begin{theorem} \lb{TA.3} Let $H_0 =-\Delta$ on $L^2 (\bbR^\nu)$. Let $V$ be a bounded function on 
$\bbR^\nu$ with $V(x)\to 0$ as $\abs{x}\to\infty$. Then 
\begin{SL} 
\item[{\rm{(i)}}] If $\nu\geq 3$ or on $L^2 ([0,\infty))$ with Dirichlet boundary conditions at $0$ 
and $V(x) \geq -\f{(\nu-2)^2}{4}\abs{x}^{-2}$, then $-\Delta + V$ has no negative spectrum. If this 
holds for all $\abs{x}>R_0$, then $-\Delta +V$ has finite negative spectrum. In any event, one has 
Hardy's inequality, 
\begin{equation} \lb{A.3} 
\f{(\nu-2)^2}{4} \int \f{\abs{\varphi(x)}^2}{\abs{x}^2} \, d^\nu x \leq \int \abs{\nabla\varphi(x)}^2 
\, d^\nu x
\end{equation}

\item[{\rm{(ii)}}] If $\nu =2$ and $V(x) \geq -\f14 (\abs{x}\log \abs{x})^{-2}$ for all $\abs{x}\geq R_0$, 
then $-\Delta +V$ has finite negative spectrum. 

\item[{\rm{(iii)}}] If $\nu\neq 2$ and $V(x) \leq - (1+\veps) \f{(\nu-2)^2}{4} \abs{x}^{-2}$ for $\abs{x} 
\geq R_0$, then $-\Delta + V$ has infinite negative spectrum. 

\item[{\rm{(iv)}}] If $\nu=2$ and $V(x)\leq - (1+\veps) \f14 (\abs{x}\log \abs{x})^{-2}$ for all 
$\abs{x} >R_0$, then $-\Delta +V$ has an infinite negative spectrum.  
\end{SL} 
\end{theorem} 

\begin{proof} By the min-max principle, it suffices to consider the case where $V$ is spherically 
symmetric. In that case, $-\Delta +V$ is unitarily equivalent (see \cite[pp.~160--161]{RS2}) to a 
discrete sum $\oplus H_{\ell,m}$ on $\oplus L^2 ([0,\infty),dr)$ where 
\[
H_{\ell,m} = -\f{d^2}{dr^2} + \f{(\nu-1)(\nu-3)}{4} \, \f{1}{r^2} + \f{\kappa_\ell}{r^2} + V 
\]
where $\kappa_{\ell=0}=0$ and all other $\kappa$'s have $\kappa_\ell >0$ and $\kappa_\ell \to\infty$. 
Since $\f{(\nu-1)(\nu-3)}{4}- \f14 = \f{(\nu-2)^2}{4}$, this result follows from the previous two 
theorems. 
\end{proof}

The following result seems to be new:

\begin{theorem} \lb{TA.4} Let $\nu\geq 3$. Let $V(x)$ be a function on $\bbR^\nu$ so that for 
any $\alpha$, $m(\alpha) \equiv \abs{\{x\mid\, \abs{V(x)}>\alpha\}}$ is finite. Suppose 
\begin{equation} \lb{A.4} 
\lim_{\alpha\downarrow 0}\, \alpha^{\nu/2} m(\alpha) < \tau_\nu \biggl( \f{\nu -2}{2} \biggr)^\nu
\end{equation}
where $\tau_\nu$ is the volume of the unit ball in $\bbR^\nu$. Then $-\Delta + V$ has a finite number 
of bound states. 
\end{theorem} 

\begin{proof} Let $V_0 = \f{(\nu-2)^2}{4}\, \f{1}{\abs{x}^2}$. 
$m_0(\alpha) \equiv \abs{\{x\mid V_0(x) > \alpha\}} =\tau_\nu (\f{\nu -2}{2})^\nu \alpha^{-\nu/2}$ so 
\eqref{A.4} implies $V=V_1 + V_2$ where $V_1\in L^{\nu/2}$ and $V_2$ has a spherical rearrangement, 
$V_2^*$ (see \cite{LLbk}) with 
\[
V_2^* \leq (1-\veps) V_0^*
\]
for some $\veps>0$. Now 
$-\Delta + V=\veps (-\Delta) +V_1 + (1-\veps) [-\Delta + (1-\veps)^{-1} V_2]$. By the 
Cwikel-Lieb-Rozenblum \cite{Cw,L1,Roz} bound, $\veps (-\Delta) + V_1$ has finite negative spectrum. 
By \eqref{A.3}, 
\[
\|V_0^{1/2}(-\Delta)^{-1} V_0^{1/2}\|\leq 1
\]
The Brascamp-Lieb-Luttinger inequality \cite{BLL} shows
\[
\|\abs{V_2}^{1/2} (-\Delta)^{-1} \abs{V_2}^{1/2} \| \leq \|V_2^{* 1/2} (-\Delta) V_2^{* 1/2}\|
\]
It follows that $-\Delta + (1-\veps)^{-1} V_2$ has no negative spectrum. 
\end{proof} 

We note that the main part of this paper has results that extend some of these results to Schr\"odinger 
operators with oscillatory potentials; see Theorem~\ref{T2.2}. We now turn to the discrete Jacobi case, 
beginning with 

\begin{theorem} \lb{TA.4x} Let $J_0$ be the free Jacobi matrix. Then 
\begin{equation} \lb{A.5} 
\sum_{n=1}^\infty \f{1}{4n^2} \, \abs{u(n)}^2 \leq (u, (2-J_0)u)
\end{equation}
\end{theorem} 

{\it Remarks.} 1. We will see below that $\f14$ is the optimal constant in this inequality, that is, 
it is false if $\f14$ is replaced by a larger constant. 

\smallskip
2. However, $\f{1}{4n^2}$ can be replaced by $\f{1}{4n^2} + \f{5}{32n^4}$ or, more generally, 
$[(1+\f{1}{n})^{1/2} + (1-\f{1}{n})^{1/2} -2]$. 

\begin{proof} There is a Sturm theory in the discrete case \cite{PF,Tes}. One needs to look at zeros 
of the linear interpolation of $u$. In particular, if $b_n$ is such that there is a positive solution 
$u_0$ of 
\begin{equation} \lb{A.6} 
(J_0 + b)u_0 =2u_0
\end{equation}
then $(2-J_0 -b)\geq 0$. Let $u_0 (n) = n^{1/2}$ for $n\geq 0$. Define for $n\geq 1$, 
\[
b_n = \f{u_0 (n+1) + u_0 (n-1)}{u_0 (n)} -2 = \biggl( 1+ \f{1}{n} \biggr)^{1/2} + 
\biggl( 1 - \f{1}{n} \biggr)^{1/2} -2 
\]
Thus \eqref{A.6} is obeyed so $(2-J_0 -b) \geq 0$ or $\sum b_n \abs{u(n)}^2 \leq (u, (2-J_0)u)$ 
for any $u$. Since $(1-x)^{1/2}= 1-\sum_{n=1}^\infty c_n x^n$ with $c_n \geq 0$ and $c_2 = \f18$, 
$b_n \geq \f{1}{4n^2}$. 
\end{proof} 

\begin{theorem} \lb{TA.5} Let $J$ be a Jacobi matrix with 
\begin{gather} 
\limsup n^2 \abs{a_n -1} =\gamma_a \lb{A.7} \\
\limsup n^2 \abs{b_n} = \gamma_b \lb{A.8}
\end{gather}
both finite with 
\begin{equation} \lb{A.9} 
2\gamma_a+ \gamma_b < \tfrac14
\end{equation}
Then $J$ has finitely many bound states outside $[-2,2]$. 
\end{theorem} 

{\it Remarks.} 1. As we will see, the $\f14$ in \eqref{A.9} cannot be improved. 

\smallskip 
2. In \cite{Chi}, Chihara proves $J$ has finitely many eigenvalues if 
\begin{equation} \lb{A.9x} 
\limsup \bigl( n^2 [(a_n^2 -1) \pm \tfrac12\, (b_n + b_{n-1})]\bigr) < \tfrac{1}{16}
\end{equation}
(We take his Jacobi matrix and multiply  by $2$ to get the $[-2,2]$ rather than $[-1,1]$ 
normalization; then his $c_n$ and $\lambda_n$ are related to ours by $a_n = \sqrt{4\lambda_{n+1}}$, 
$b_n =2c_n$.) This leads to $2\gamma_a + \gamma_b <\f{1}{16}$, so our result, which is best possible, 
is better by a factor of $4$. 

\smallskip
3. Because having no eigenvalues remains true if the $a_n$'s are decreased, \eqref{A.7} can be 
replaced by $n^2 (a_n -1)_+$, although it still must be true that $a_n\to 1$ as $n\to\infty$. 

\begin{proof} By \eqref{4.1}, it suffices to prove the spectrum above $2$ is finite. Pick $\veps$ 
so that 
\begin{equation} \lb{A.10} 
\f{2\gamma_a + \gamma_b + 3\veps}{1-\veps} \leq \f14 
\end{equation}
By changing $a_n$ and $b_n$ on a finite set (which, because it is a finite rank perturbation of $J$, 
cannot change the finiteness of the number of eigenvalues), we can assume for all $n$, 
\begin{equation} \lb{A.11} 
\abs{a_{n-1} -1} + \abs{a_n -1} \leq \f{2(\gamma_a + \veps)}{n^2} \qquad a_n -1\geq -\veps  
\qquad \abs{b_n} \leq \f{\gamma_b +\veps}{n^2}
\end{equation}

By \eqref{5.1}, we then have 
\begin{align*}
(u, (2-J)u) &\geq (1-\veps) (u, (2-J_0)u) - \sum_{n=1}^\infty \f{2\gamma_a + \gamma_b + 3\veps}{n^2} 
\, u_n^2 \\
&\geq (1-\veps) \biggl[(u, (2-J_0)u) - \tfrac14 \sum_{n=1}^\infty \f{u_n^2}{n^2} \biggr] \geq 0 
\end{align*}
where we first use \eqref{A.10} and then \eqref{A.5}. 
\end{proof} 

In the other direction, we have 

\begin{theorem} \lb{TA.6} Let $J$ be a Jacobi matrix with 
\begin{gather} 
\liminf n^2 (a_n -1) = \gamma_a \lb{A.12} \\
\liminf n^2 b_n = \gamma_b \lb{A.13} 
\end{gather} 
with $\gamma_a \geq 0$, $\gamma_b \geq 0$, and 
\begin{equation} \lb{A.14} 
2\gamma_a + \gamma_b > \tfrac14
\end{equation}
Suppose also that 
\[
\lim_{n\to\infty} \, \abs{a_n -1} + \abs{b_n} =0
\]
Then $J$ has an infinity of eigenvalues in $[2,\infty)$. 
\end{theorem} 

{\it Remark.} The existence of $O(\f{1}{n^2})$ potentials with an infinity of eigenvalues evoked some 
interest because Case \cite{Case} claimed that if $\sup n^2 [\abs{a_n-1} + \abs{b_n}]<\infty$, there 
were only finitely many eigenvalues. Chihara \cite{Chi} produced a counterexample with $b_n \sim 
\f{1}{2n^2}$, $a_n -1 \sim \f{3}{8n^2}$ (after changing to our normalization), so $2\gamma_a + \gamma_b 
= \f54$, larger than the needed $\f14$ our theorem allows. 

\begin{proof} If there are only finitely many eigenvalues, the solution of $(J-2)u=0$ with $u(0)=0$ 
has only finitely many zeros so, by restricting to the region beyond the zeros and using Sturm theory, 
we see there is an $N_0$ so 
\begin{equation} \lb{A.15} 
u(n)=0, \ n\leq N_0 \Rightarrow \langle u, (2-J)u\rangle \geq 0
\end{equation}
Define $\tilde a_n =\min (a_n, 1+\f{\gamma_a}{n^2})$, $\tilde b_n =\min (b_n, \f{\gamma_b}{n^2})$ so 
\begin{equation} \lb{A.16} 
\lim [n^2 (\tilde a_n -1) + n^2 (\tilde a_{n-1}-1) + n^2 \tilde b_n] = 2\gamma_a + \gamma_b > 
\tfrac14
\end{equation} 
By \eqref{5.1} and \eqref{A.15} if $u(n)=0$ for $n\leq N_0$, 
\begin{align}
0 & \leq (u, (2-J)u)
\notag \\ 
&= 
\sum_{n=1}^\infty a_n (u(n) - u(n+1))^2 + \sum_{n=1}^\infty (-b_n - (a_n-1) - (a_{n-1} -1))u(n)^2 
\notag \\
&\leq 
\sum_{n=1}^\infty a_n (u(n) - u(n+1))^2 + \sum_{n=1}^\infty (-\tilde b_n - (\tilde a_n -1) 
- (\tilde a_{n-1}-1)] u(n)^2 \lb{A.17} 
\end{align} 
since $b_n \geq \tilde b_n$, etc. 

Given $\varphi\in C_0^\infty (0,\infty)$ and $\ell=1,2,\dots$, pick 
\begin{equation} \lb{A.18} 
u_n^\ell = \sqrt{\ell}\, \varphi \biggl( \f{n}{\ell}\biggr) 
\end{equation}
Since $\supp(\varphi)$ is a compact subset of $(0,\infty)$, $u_n^{(\ell)}=0$ if $n\leq \veps\ell$ for 
some $\veps >0$, so \eqref{A.17} holds for $\ell$ large. Since $a_n\to 1$, 
\begin{align*} 
\sum_{n=1}^\infty a_n (u_n^{(\ell)} - u_{n+1}^{(\ell)})^2 &= \sum_{n=1}^\infty a_n 
\biggl( \f{\varphi(\f{n}{\ell}) - \varphi (\f{n+1}{\ell})}{\f{1}{\ell}}\biggr)^2 \f{1}{\ell} \\
&\to \int \varphi'(x)^2\, dx 
\end{align*}
Similarly, by \eqref{A.16}, 
\begin{align*}
\sum_{n=1}^\infty [-\tilde b_n - &(\tilde a_n -1) - (\tilde a_{n-1}-1)] u_n^2 \\
&= \sum_{n=1}^\infty n^2 [-\tilde b_n - (\tilde a_n -1) - (\tilde a_{n-1}-1)] \, 
\f{\varphi (\f{n}{\ell})^2}{(\f{n}{\ell})^2} \, \f{1}{\ell} \\
&\to - \int \biggl( \f{2\gamma_a + \gamma_b}{x^2} \biggr) \varphi(x)^2\, dx
\end{align*}
we thus have that 
\[
-\f{d^2}{dx^2} - \f{2\gamma_a + \gamma_b}{x^2} \geq 0
\]
violating Theorem~\ref{TA.1}(ii). This contradiction proves that $J$ must have infinitely many 
eigenvalues. 
\end{proof}

\bigskip


\end{document}